\documentclass[prb,showpacs,byrevtex,twocolumn]{revtex4-1}
\usepackage{graphicx}
\usepackage{amsmath} 
\usepackage{amsfonts} 
\usepackage[section]{placeins}
\usepackage{color}

\begin{document}

\title{Factorization, coherence and asymmetry in the  Heisenberg spin-1/2 XXZ chain a transverse magnetic field and with Dzyaloshinskii-Moriya interaction}

\author{Pradeep Thakur, P. Durganandini}

\affiliation{Department of Physics, Savitribai Phule Pune University, Pune-411007, INDIA}

\date{\today}

\begin{abstract}
We investigate  the factorization, coherence and asymmetry  properties  of the one dimensional Heisenberg spin-1/2 XXZ chain with Dzyaloshinskii-Moriya interaction (DMI) and a transverse magnetic field using quantum information measures. Both longitudinal and transverse DM vectors are considered. Using  numerical DMRG methods, we compute bipartite entanglement estimators like the one-tangle, two-spin concurrence and quantum coherence estimators like the Wigner-Yananse-skew information.  We show that a  longitudinal DMI destroys the factorizability property while a  transverse DMI preserves it. We relate the absence of factorizability to the breaking of the $U(1)$ rotation symmetry about the local magnetization axis at each lattice site. Physically, the breaking of the symmetry manifests in the existence of a chiral current.  Further, we show that although the longitudinal DMI destroys the factorization property, there is a `pseudofactorizing' field at which the entanglement and hence violation of the $U(1)$ symmetry is minimal.  Our calculations indicate a phase coherent ground state at $h_{pf}$.  An entanglement transition (ET) occurs across this field which is characterized by an enhanced but finite range of two-spin concurrence in its vicinity in contrast with the diverging range of the concurrence for the ET across the factorizing field.  We relate the asymmetry to the `frameness' or the ability for the state to act as a reference frame for some measurement. In the absence of the longitudinal DMI (or in the presence of a transverse DMI),  at the factorizing field,  the single site magnetization axis serves to specify the common $z$-axis  for the full system but not the full Cartesian reference frame due to a lack of phase reference.  On the other hand, in the presence of a longitudinal DMI,  our results indicate that at the pseudofactorizing field, the local magnetization and the chiral current are sufficient to specify the full  Cartesian reference frame with the  chiral current serving as the macroscopic quantity to determine the phase reference.

\end{abstract}

\maketitle

\section{Introduction}

    An interesting and intriguing feature of a certain class of quantum phase transitions(QPT) is that they are associated with the existence of a non-trivial `factorizability'  property,  i.e., the quantum state becomes completely separable at certain parameter strengths~\cite{kurmann}. The existence of such factorizing points serves as a precursor signalling the existence of a QPT associated with an entanglement transition(ET); there is a crossover from one type of entanglement to another across the factorizing field. ET have been characterized by the divergence of the range of pairwise entanglement close to the factorization point~\cite{roscilde2004, fubini2006, amicoRange2006}.  The most notable example is that of the Heisenberg spin $S= 1/2 $ chain in an external magnetic field where it was shown several years ago~\cite{kurmann} that a  factorizable ground state emerges at a certain value of the magnetic field. Further impetus in the subject has been provided  by the use of quantum information measures to elucidate the conditions for the existence and location of such factorizable ground states~\cite{giampaoloPRL}.  This has led to several studies of the effects of additional interactions and generalizations to higher spins on the critical and factorizability properties of the system~\cite{giampaolo2009, canosa2010, cerezo2015, giampaolo2018}.  In this context, a particularly interesting additional interaction to consider is the Dzyaloshinskii-Moriya interaction(DMI) which is of the form: $\vec D. (\vec{S}_n \times\vec{S}_{n+1})$ . The DMI, while originally shown to arise from spin-orbit interaction~\cite{DM1, *DM2},  occurs in a variety of physical contexts~\cite{perk, satoPRL, mf-review, knb}, and has been shown to lead to several unusual effects like the field induced gap in copper benzoate compounds~\cite{oshikawa-affleck1},  various field induced phases,  chiral order, magnetoelectric effects, etc.~\cite{garate-affleck, starykh, brockmann,pradeepAIP,pradeepPRB} The presence of DMI also modifes the entanglement and quantum correlation properties~\cite{ derzhkoetal, jafarietal, soltani, radhaDMI2017, radhaSCIREP2017, yi2019}.     
    
  In  earlier work~\cite{pradeepAIP, pradeepPRB}, we had studied the effect of a longitudinal DMI $D_z$(which we interpreted there as an electric field ) in the  anisotropic Heisenberg spin $S=1/2$  $XXZ$ model and showed that it does not lead to any new phases; it only modifies the phase boundaries; increasing the disordered phase.   An interesting question to ask is whether the DMI modifies the entanglement properties of the system and in particular, whether  the factorizability phenomenon is preserved in the presence of  a DMI.   While there are some recent studies probing the effect of DMI on the  factorization and critical properties of the spin chains using quantum coherence measures~\cite{soltani, radhaDMI2017, radhaSCIREP2017, yi2019}, most of these studies have either focussed on the analytically solvable $XY$ model which can be mapped into the free fermion model or used perturbative methods to solve the interacting fermion model.  In this work,  we address the question of the effect of the DMI on the factorization, coherence and asymmetry  properties  of the one dimensional spin-1/2 anisotropic Heisenberg XXZ model in a transverse magnetic field by analyzing the quantum information measures of entanglement and coherence.   Specifically, we analyze the entanglement properties by computing bipartite entanglement and coherence estimators like the one-tangle~\cite{roscilde2004}, two-spin concurrence~\cite{wootters1998}, the Wigner-Yanase-skew information(WYSI)~\cite{wigner1963}, etc by using numerical DMRG techniques. Both longitudinal and transverse DM vectors are considered. 

 Our main result is that in the presence of the DMI, the critical and factorizability properties  crucially depend on both the orientation and magnitude of the DMI.  A  transverse DMI (below a certain critical strength) preserves the factorizability property in the AFM phase. On the other hand, even a small  longitudinal DMI destroys the factorizability property.  We attribute the difference in the two cases to the breaking of a local $U(1)$ symmetry.  In the absence of the DMI and at the factorization point,  the full many body state can be described by just one single parameter: the  on-site magnetization. There is a local $U(1)$ rotation symmetry of the ground state about the magnetization axis. An additional transverse DMI preserves the local $U(1)$ rotation symmetry, while the longitudinal DMI breaks the  symmetry.  Physically, the breaking  of the $U(1)$ symmetry manifests in the existence of a chiral current in the antiferromagnetic phase in the presence of a longitudinal DMI.  Further, we show that although the factorizability property is lost in the presence of a longitudinal DMI,  there exists a `pseudofactorizing'(PF) magnetic field at which the violation of the $U(1)$ symmetry is minimal and the ground  state can be described by a macroscopic phase coherent wave function with minimum bipartite entanglement and extremal coherence.  We also identify an ET across the PF  field characterized by an enhanced but finite range of pairwise entanglement in the vicinity of the PF field.  We relate the asymmetry to the `frameness'~\cite{bartlett, gour}. In the absence of the longitudinal DMI, at the factorizing field,  the single site magnetization axis serves to specify the common $z$-axis  for the full system but not the full Cartesian reference frame due to a lack of phase reference.  On the other hand, in the presence of a longitudinal DMI,  our results indicate that at the pseudofactorizing field, the local magnetization and the chiral current are sufficient to specify the full  Cartesian reference frame with the  chiral current serving as the macroscopic quantity to determine the phase reference.
 
                The paper is organized as follows:  we begin in Sec.\ref{sec:DMI} by  presenting the results for the effect of a transverse DMI on the ground state properties of the system using numerical DMRG methods. We obtain the ground state phase diagram by computing various  ground state quantities like the energy gaps, magnetization, spin currents, etc. By computing the bipartite entanglement estimators, namely the one tangle and two spin concurrence, we show that the transverse  DMI preserves the factorizability property.  We then discuss the corresponding results for the case of a longitudinal DMI and show that it destroys the factorizability property. The nature of the entanglement in the different phases and associated entanglement transitions is described in Sec.~\ref{sec:ET}. The quantum coherence and symmetry properties in the presence of DMI are described  in Sec.~\ref{sec:coherence}. Finally, we conclude with a brief summary and discussion of our results in Sec.\ref{sec:concl}.

   \section{\label{sec:DMI} XXZ chain in the presence of DMI}
        The anisotropic Heisenberg spin $1/2$ XXZ-chain in the presence of magnetic fields and the Dzyaloshinskii-Moriya interaction is described by the Hamiltonian:
\begin{eqnarray}
\mathcal{H} = {\sum_{i=1}^N} [J(S_i^x S_{i+1}^x + S_i^y S_{i+1}^y + \Delta S_i^z S_{i+1}^z) \nonumber \\
+ \vec{D}\cdot(\vec{S}_i \times\vec{S}_{i+1}) - \vec h \cdot \vec{S}_i] \label{eq:ham_dm}
\end{eqnarray}
where $S_i^a$, with $a = x, y, z$,  describe the components of the spin $1/2$ operator at the $i$-th site along the chain, $\Delta$ is the easy-axis anisotropy (the $xy$-plane being the easy-plane), $ \vec h$ denotes the external magnetic field while $ \vec D $,  the DM vector couples to the chirality operator: $\vec{K} \equiv  \vec S_{i} \times \vec S_{i+1}$.  
In the absence of the magnetic field and DMI, and for large Ising anisotropy, the XXZ model has an antiferromagnetic(AFM) ground state.  External magnetic fields modify the ground state behaviour depending on the strength and direction of the field.  Longitudinal magnetic fields directed along the $z$ direction disorder the AFM order at a certain critical field $h_{c1}$ leading to a critical gapless regime  for $h_{c1} < h_z < h_{c2}$. At critical field strength $h_{c2}$, there is a transition from the critical gapless phase to a gapped saturated ferromagnetic (FM) behaviour leading to a gapped  regime with FM order for field strengths $h_z > h_{c2}$.  At the critical field $h_{c2}$, where the transition from the critical gapless phase to the gapped saturated ferromagnetic (FM) phase occurs, all quantum correlations get suppressed  and the ground state becomes the classical fully separable ferromagnetic state.

         Transverse magnetic fields on the other hand cause a phase transition from an AFM ordered phase to a FM polarized phase at a critical field strength $h_x=h_{cr}$ with the field induced magnetization saturating only as $h_x\rightarrow \infty$.  While one would expect then  that all quantum correlations get suppressed only as $h_x \rightarrow \infty$, remarkably,  due to an intricate balancing between the exchange interactions and the external field, there exists an intermediate non-trivial field strength $h_f (<h_{cr})$ within the AFM phase, at which the ground state becomes a classical, fully separable factorized state~\cite{kurmann}.  Further, it has been shown that the existence of such a factorization field signals  a so-called entanglement transition where the entanglement changes between parallel and antiparallel types in the ground state concurrence and is characterized by the divergence of the range of pairwise entanglement close to the factorization point ~\cite{roscilde2004}.
In  earlier work~\cite{pradeepAIP, pradeepPRB}, we had studied the effect of a longitudinal DMI $D_z$(which we interpreted there as an electric field ) in the anisotropic  $XXZ$ model and showed that it does not lead to any new phases; it only modifies the phase boundaries; increasing the disordered phase. For $D_z$ smaller than a certain critical strength $D_c$, there are two gapped phases: an antiferromagnetically ordered phase ($AFM_z$) for $h_x<h_{cr}$ and a gapped ferromagnetic ($FM_x$) phase for $h_x>h_{cr}$. The $AFM_z$ phase corresponds to a phase with a staggered magnetization $M^z_s$ along the $z$-direction and  a uniform magnetization $M^x$ along the $x$-direction~\cite{pradeepPRB}.  There is also a finite chiral current $\langle K^z \rangle$ (termed as electric polarization $P^y$ in Ref.~\cite{pradeepPRB}) in this phase. The transition to the $FM_x$ phase occurs at a critical transverse field strength $h_x=h_{cr}$, the value depending on the strength of $D_z$.  In this work, we study the interplay of the effect of a transverse magnetic field and DMI on the quantum correlation properties by analyzing various bipartite quantum correlation measures, specifically, the one-tangle, two spin concurrence and the WYSI. In particular, we investigate the
 question of the existence of the ground state factorizability phenomenon in the presence of a DMI . We consider both  longitudinal and transverse DMI.  

 The bipartite quantum correlation
measures can be obtained from the two-spin reduced density matrix $\rho^{(2)}_{ij}$ by tracing out from the full ground state density matrix, all the spins except those at the lattice sites $i$ and $j$. It can be expressed most generally in terms of the various two spin  correlation functions as:
\begin{align}
\rho^{(2)}_{ij} = \frac{1}{4}
\begin{pmatrix}
a_+ & e_+& h_+ & f_- \\
e^*_+ & a_- & f_+ & h_- \\
h^*_+ & f_+^* & c_- & e_- \\
f_-^* & h^*_- & e^*_- & c_+ \\
\end{pmatrix} \nonumber 
\end{align}
where ($\sigma _i^a = 2 S_i^a$),
\begin{align}
&a_{\pm} = 1 + (\langle \sigma_i^z \rangle \pm \langle \sigma_j^z \rangle) \pm  \langle \sigma_i^z \sigma_j^z \rangle \nonumber \\
&c_{\pm} = 1 - (\langle \sigma_i^z \rangle \pm \langle \sigma_j^z \rangle) \pm \langle \sigma_i^z \sigma_j^z \rangle \nonumber \\
&e_{\pm} = \langle \sigma_j^x \rangle - i\langle \sigma_j^y \rangle \pm( \langle  \sigma_i^z \sigma_j^x \rangle - i\langle \sigma_i^z \sigma_j^y \rangle ) \nonumber \\
&f_{\pm} = \langle \sigma_i^x\sigma_j^x \rangle \pm  \langle \sigma_i^y\sigma_j^y \rangle \pm i ( \langle \sigma_i^x\sigma_j^y \rangle \mp \langle \sigma_i^y\sigma_j^x \rangle ) \nonumber \\
&h_{\pm} = \langle \sigma_i^x \rangle - i\langle \sigma_i^y \rangle  \pm (\langle  \sigma_i^x \sigma_j^z \rangle - i\langle \sigma_i^y \sigma_j^z \rangle )\\
\label{eqn:2rho}
\end{align}
The one-spin reduced density matrix  $\rho_i^{(1)}$ at lattice site $i$ can be then obtained by tracing out, say, the second spin from the above $\rho^{(2)}$ as:
\begin{align}
\rho_i^{(1)} = 
\begin{pmatrix}
a_++a_- & h_+ + h_-  \\
h^*_+ + h^*_- & c_+ + c_- \\
\end{pmatrix}
\end{align}
The one-tangle in the spin systems we consider is defined as an entropic measure of the bipartite entanglement between a single spin , say, at the $i^{th}$ site and the rest of the spins and can be obtained  in terms of  the one-spin reduced density matrix $\rho_i^{(1)}$ as: 
 \begin{equation}
 \tau_i=4 \cdot \text{det} \rho_i^{(1)} =  1 - 4 \langle \vec S_i\rangle ^2 
 \end{equation}
  It represents a global estimate of the entanglement in a translationally invariant system since it does not depend on the site ( we therefore drop in the following, the site index in $\tau$). It  has been shown that the vanishing of the one-tangle is a necessary and sufficient condition for the existence of a factorized ground state in a translationally invariant system~\cite{giampaolo2007}.  It is most useful when the state of the system is pure; for a mixed state, the one-tangle actually gives us an upper bound on the amount of entanglement. The two spin concurrence $C_{i,i+n}$ quantifying  the entanglement of a pair of spins is defined as~\cite{hill1997}:
\begin{equation}
C(\rho) = \text{max}\{ 0,\lambda_1-\lambda_2-\lambda_3-\lambda_4\},
\end{equation}
where the $\lambda_i (\geq 0) $ are the square roots in decreasing order of the eigenvalues of the non-Hermitian matrix $R=\rho^{(2)}_{ij} \tilde{\rho}^{(2)}_{ij}$ where 
 $ \tilde{\rho} = (\sigma_y \otimes \sigma_y) \rho^{*} (\sigma_y \otimes \sigma_y)$ is the  spin-flipped density matrix corresponding to the two-spin density matrix $ \rho ^{(2)}$~\cite{hill1997, wootters1998}.   
   
\subsection{\label{sec:Dx}Effect of  a transverse DMI}

\begin{figure}[!htbp]
\includegraphics[width=3.0in]{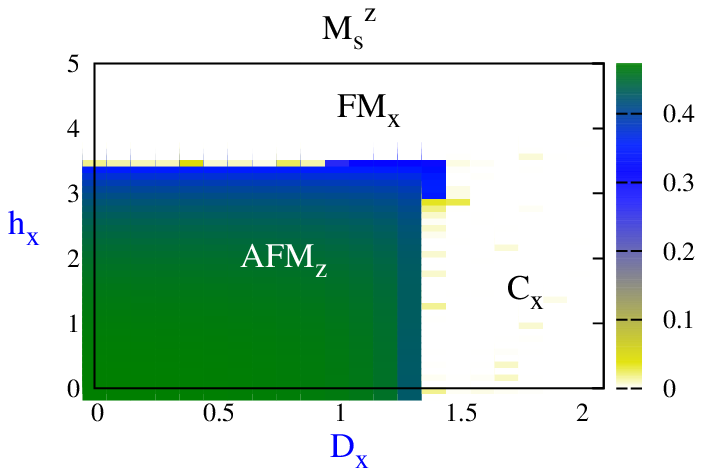}{(a)}
\includegraphics[width=3.0in]{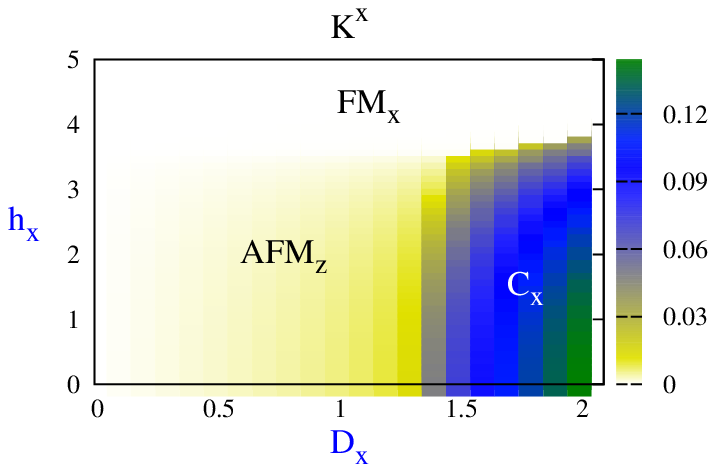}{(b)}
\includegraphics[width=3.0in]{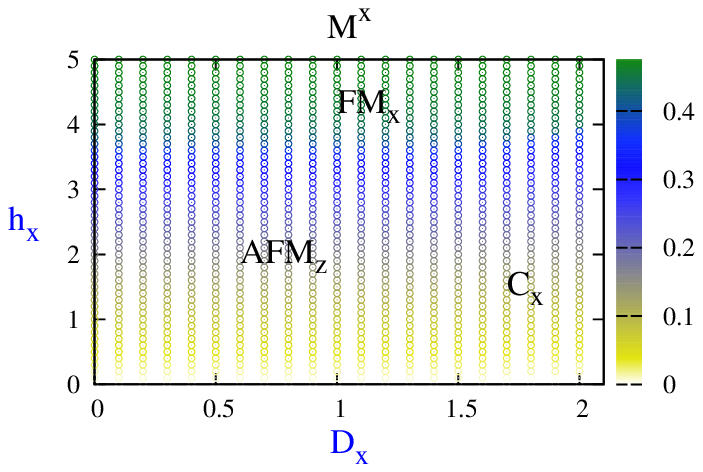}{(c)}
\includegraphics[width=3.0in]{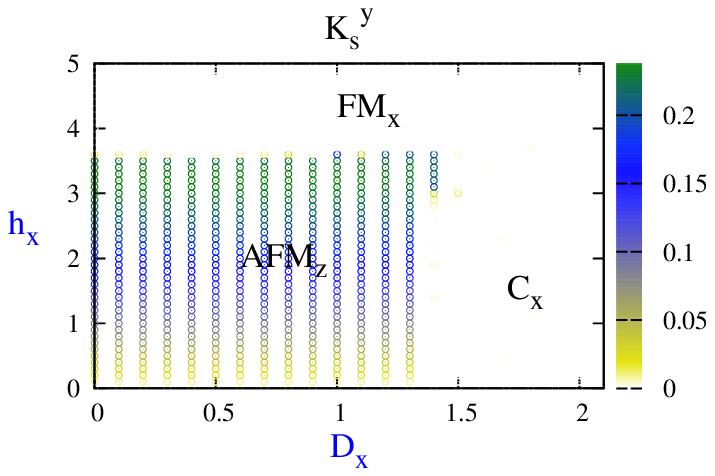}{(d)}
\caption{
The $h_x-D_x$ dependence of the  (a) staggered $z$-magnetization per site, $M_s^z$, (b)  uniform chiral order per bond in the $x$ direction, $K^x$ , (c) uniform $x$-magnetization per site, $M^x$, (d) staggered chiral order $K^y_s$ per bond .
The DMRG computations were performed on a 64 site chain with open boundary conditions for  anisotropy $\Delta =4.5$.  The truncation error in the DMRG program was set to $10^{-6}$, with $10$ sweeps. We restricted the maximum number of allowed states for a block to $70$. 
}
\label{fig:order_hxDx}
\end{figure}

 We begin with the case where both the magnetic field and the DM vector  are pointing along the $x$ axis; or  a transverse magnetic field, $h_x$, and a transverse DMI, $D_x$ ($D_z=0$).  We compute numerically (using the ALPS DMRG application~\cite{ALPS-2}), the energy gaps, various physical observables like magnetization and chiral currents and also the bipartite entanglement measures, namely the one tangle and two spin concurrence,  in order to identify and characterize  the different phases.
 The behaviour of the staggered magnetization ,$M_s^z$, the uniform chiral order $K^x$ ,  uniform magnetization $M^x$  and the staggered chiral order $K_s^y$ as a function of $h_x-D_x$  are shown in Fig.~\ref{fig:order_hxDx}(a-d). 
Depending on the relative strengths of the transverse magnetic field and the DM field,  three distinct phases can be identified as shown in Fig.~\ref{fig:order_hxDx}: \\

\noindent (i)  Staggered chiral antiferromagnetic $AFM_z$ phase: \\
For field strengths $h_x<h_c$, $D_x<D_c$, there is a gapped phase with  near-saturation antiferromagnetic magnetic order along the $z$ direction, unsaturated $M^x$ and a staggered chiral order $K_s^y$ along the $y$ direction as shown in Fig.~\ref{fig:order_hxDx}(a-d)). The staggered chiral order $K_s^y$ is induced by the transverse magnetic field and for small $h_x$, is  linearly proportional to $h_x$. \\

\noindent (ii) Field induced Ferromagnetic $FM_x$ phase: \\
For transverse magnetic fields larger than the critical field strength $h_c$ ($h_x>h_c$ ) and $D_x <D_c$,  there is a ferromagnetic phase  with only induced saturated ferromagnetic $M^x$ order  as can be seen from panels (a) and (c) of Fig.~\ref{fig:order_hxDx}. Further, it can be seen from panels (b) and (d) of Fig.~\ref{fig:order_hxDx} that there is no chiral order in this phase.\\

\begin{figure*}[!htpb]
\includegraphics[width=3.0in]{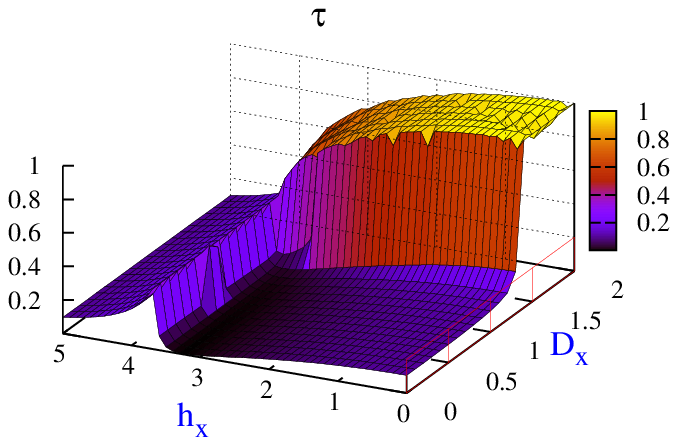}(a)\hfill
\includegraphics[width=3.0in]{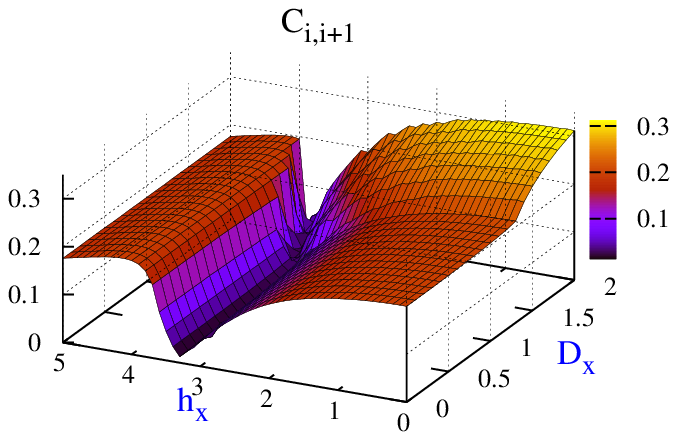}{(d)}
\\[0.1cm]
\includegraphics[width=3.0in]{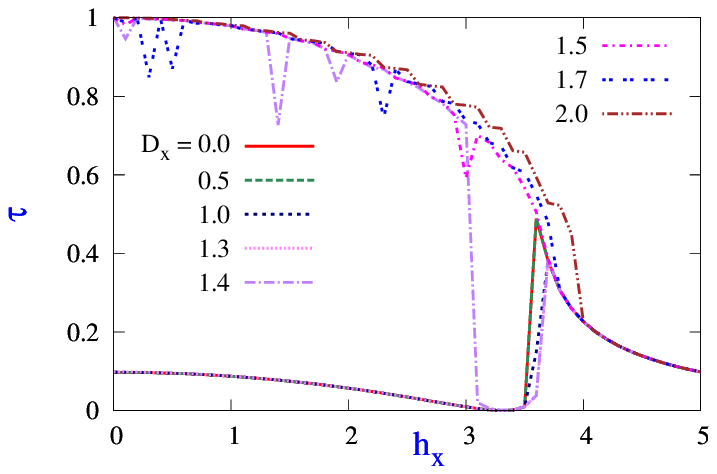}(b) \hfill
\includegraphics[width=3.0in]{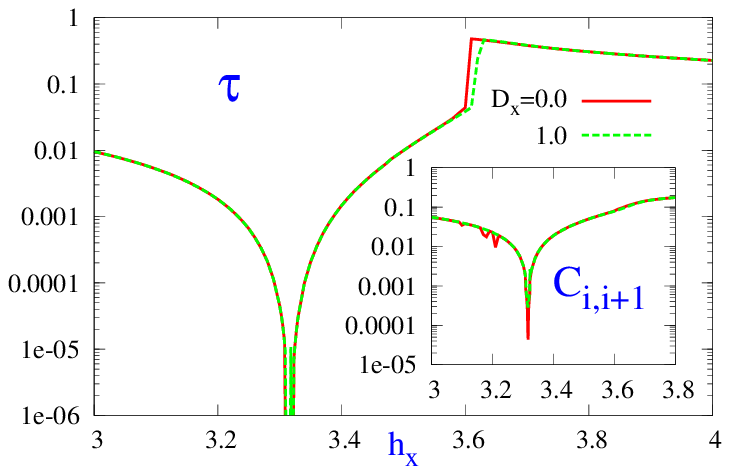}{(e)}
\\[0.1cm]
\includegraphics[width=3.0in]{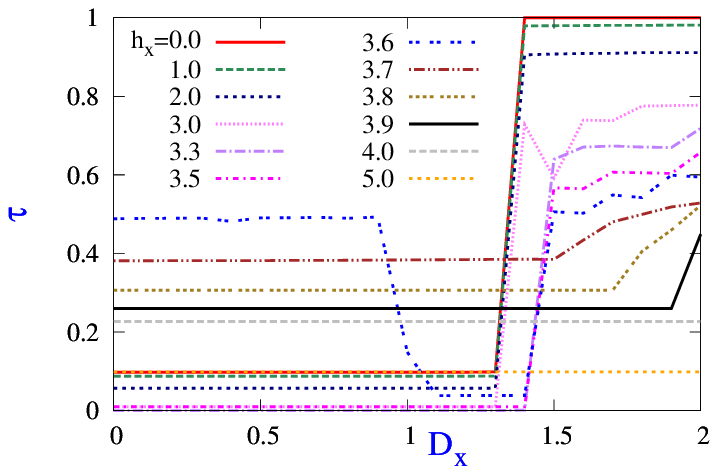}{(c)} \hfill \includegraphics[width=3.0in]{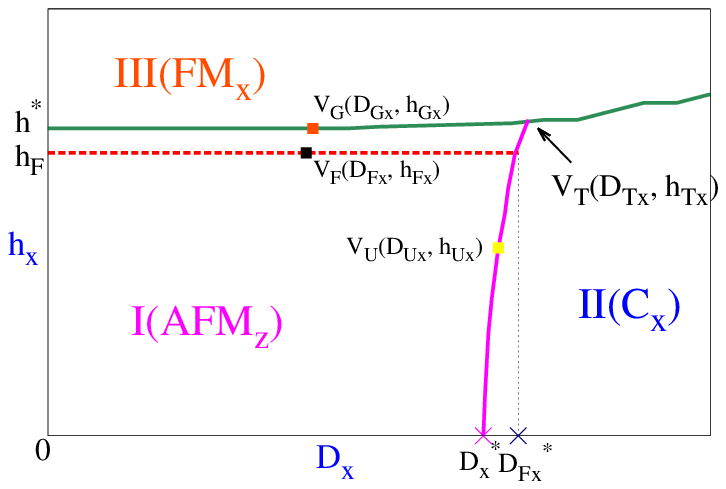}(f)
\caption{(a) The $h_x-D_x$ dependence of the one-tangle, $\tau$.  (b) The $h_x$ dependence of $\tau$ for different $D_x$ values. (c) The $D_x$ dependence of $\tau$ for different $h_x$ values.  (d) The $h_x-D_x$ dependence of the nearest neighbour concurrence. (e)  The semilog plot of the one-tangle as a function of $h_x$ in absence and presence of $D_x$. It. vanishes at $h_x=h_f \sim 3.316$ both in the presence and absence of  $D_x$. The inset shows the $h_x-D_x$ dependence of the NN concurrence, $C_{i,i+1}$.  (f)Schematic ground state phase diagram with $h_x$ and $D_x$ as parameters, obtained using DMRG  computations of energy gaps and various observables like the magnetizations, entanglement measures, etc. The factorizing curve is the dashed curve lying inside the $AFM_z$ phase. All other parameters are as in Fig.1.}
\label{fig:dx_factor}
\end{figure*}

\noindent (iii)  Chiral phase:\\
For large D-M interaction strength beyond a certain critical value $D_x^*$, ( $D_x>D_c$)  and magnetic field strengths $h_x <h_c$, there is a uniform chiral phase  with only an induced uniform chiral order $K^x$ as shown in Fig.~\ref{fig:order_hxDx}(b,d). There is no magnetic order in this phase as can be seen from Fig.~\ref{fig:order_hxDx}(a, c).
\vskip 0.2cm
The entanglement properties of the system can be analyzed from the behaviour of the  one-tangle and concurrence  in the different phases which we show in Fig.~\ref{fig:dx_factor} (a-e). The magnetic and entanglement properties have been summarized by the schematic shown in Fig. ~\ref{fig:dx_factor}(f).  The $h_x-D_x$ dependence of the one tangle is shown in Fig. \ref{fig:dx_factor}(a).  We also plot in Fig.~\ref{fig:dx_factor}(b,c), the single parameter dependences of $\tau$ in order to obtain a better understanding of the behaviour of $\tau$ in the different phases. From Fig.~\ref{fig:dx_factor}(a), we can observe that the one-tangle shows distinct behaviour in the three phases. The one-tangle $\tau$ is small in the magnetically ordered $FM_x$ and $AFM_z$ phases  while it is large and goes to a maximum($\tau=1$) in the uniform chiral phase.  In the absence of the DMI ($D_x =0$),  the one tangle has a non-monotonic $h_x$ dependence as can be seen from Fig.~\ref{fig:dx_factor}(a,b), starting from a finite value at $h_x=0$ (in our case, $\tau\sim 0.1$ at $h_x=0$),  which decreases with increase in $h_x$,  vanishes  to zero at $h_x=h_f$,  then rises sharply to a maximum at $h_x=h_{cr}$ and then slowly decreases monotonically as $h_x$ increases further.  The vanishing of the one-tangle  at $h_x=h_f$ marks the existence of a factorized or separable state at $h_x=h_f$~\cite{kurmann}  while the sharp rise in $\tau$ at $h_x=h_{cr}$ signals the transition from the $AFM_z$ phase to the $FM_x$ phase. 
Further, it can be seen from the above plots that even in the presence of $D_x$, the one-tangle has a similar non-monotonic  $h_x$ dependence for $D_x <D_x^*$, where $D_x^*$ is the value beyond which the chiral phase emerges when $h_x=0$. In general, the chiral phase emerges at $D_{Ux}$, for the corresponding value, $h_{Ux}$, of the magnetic field (see Fig. \ref{fig:dx_factor}(f)). For $D_x$ values greater than $D_{Ux}$, the one-tangle does not vanish for any $h_x$ although the $h_x$ dependence remains non-monotonic. For example, it can be seen from Fig.~\ref{fig:dx_factor}(b) that at $D_x=1.4$, $\tau$ starts from the saturated value, $\tau=1$, begins to drop sharply after the point $(h_{Ux},D_{Ux})$ (not marked), decreases to a nonzero minimum at  $h_x \approx 3.5$, rises to a sharp maximum and then decreases slowly to zero as $h_x\rightarrow\infty$, in the $FM_x$ phase.  Similar behaviour occurs for $D_x^*<D_x<D_{Tx}$, i.e. on the curve $V_U$.
As $D_x$ is further increased, for  $D_x > D_{Ux}$,  i.e. to the right of the curve $V_U$ in Fig. \ref{fig:dx_factor}(f), there are two phases: the  uniform chiral phase, with an almost saturated $\tau (\approx 1) $, and the $FM_x$ phase, with a small $\tau$. The $h_x$ dependence of $\tau$ in this region shows a kink at the transition between the uniform chiral and the $FM_x$ phases as can be seen from Fig.~\ref{fig:dx_factor}(b). From the  $D_x$ dependence(for fixed $h_x$) of the one-tangle  shown  in Fig.~\ref{fig:dx_factor}(c), we can see that for small $h_x$ values ($h_x <h_{cr}$), 
 the one tangle has almost a constant small value for $D_x<D_{Ux}$ and rises sharply to its maximal value at $D_x > D_{Ux}$  For $h_x$ values greater than $h_{cr}$,  there is a  transition between $FM_x$ phase (for $D_x <{D_{Ux}}$)  to the chiral phase (for $D_x>D_{Ux}$) with the one tangle which is almost constant in the $FM_x$ phase changing abruptly to another constant value in the chiral phase as can be seen from the plot shown in Fig.~\ref{fig:dx_factor}(c) for say, $h_x=3.9$.  
We next examine the  behaviour of the two spin  concurrence.   From Fig.~\ref{fig:dx_factor}(d), which shows the $h_x-D_x$ dependence of the nearest neighbour( NN) concurrence, we can see that the NN concurrence has a qualitatively similar behaviour to the one tangle. In the $AFM_z$ phase, there is a similar non-monotonic $h_x$ dependence as the one-tangle in the $AFM_z$ phase, vanishing identically along the factorizing curve inside the $AFM_z$ region and rising sharply in the transition region to the $FM_z$ phase.  The NN concurrence becomes large in the chiral phase. We also comment here on a  notable difference between the one-tangle and the NN concurrence:  the two quantities behave quite  differently  in  the transition region from the chiral phase to the $FM_x$ region. While the  one-tangle does not vanish for any $h_x$ outside of the $AFM_z$ region, the NN concurrence nearly vanishes near the transition from the chiral to the $FM_x$ region as can be seen from Fig.~\ref{fig:dx_factor}(d).  
A valley of minima in the NN concurrence is seen in this region, whereas in case of the one-tangle, there’s a kink at the transition, but no minimum.
As one moves in parameter-space in the direction of increasing $h_x$, one encounters a minimum in $C_{i,i+1}$, before the chiral-$FM_x$ transition. $C_{i,i+1}$ decreases monotonically to a minimum before the phase transition, then increases to a maximum in the $FM_x$ phase and then decreases asymptotically to zero. This implies that the pairwise entanglement decreases as one nears the transition to the ferromagnetic phase in the $h_x-D_x$ parameter space.  However, one does not observe any minimum in the one-tangle near the chiral-$FM_x$ phase transition. The one-tangle is large in the chiral phase, and decreases as one moves towards its boundary with the $FM_x$ phase with a kink in the one-tangle at the transition. In particular, both the one tangle and two spin concurrence vanish at the factorizing field as can be seen from the semilog plots of the $h_x$ dependence of the one-tangle and NN two spin concurrence for representative $D_x$ values shown in panel (e) of Fig.~\ref{fig:dx_factor}.  Thus the one-tangle and the two spin concurrence characterize distinct entanglement behaviour in the three phases and three transition regions which we have summarized by the schematic in Fig. ~\ref{fig:dx_factor}(f).  The magnetically ordered phases correspond to regions with small $\tau$ and NN concurrence while the magnetically disordered chiral phase corresponds to a phase with maximal $\tau$ and large NN concurrence.  In the vicinity of the sharp peak representing the $AFM_z-FM_x$ transition, there is a `factorizing' curve inside the $AFM_z$ phase (marked as the dashed red curve in the $AFM_z$ phase in the schematic shown in Fig. ~\ref{fig:dx_factor}(f)), along which the one-tangle vanishes ($\tau =0$); the one tangle rises sharply to a maximum at $h_x =h_{cr}$ along the critical curve (marked in the schematic as the green curve). The NN concurrence also vanishes identically along the factorizing curve.  The occurrence of the factorizing  line and the sharp peak in its vicinity can be used to identify the  $AFM_z$-$FM_x$ transition, and thus the two magnetically ordered phases - $AFM_z$ and $FM_x$.  The chiral phase has a  maximal $\tau$ value which is characteristic of a phase possessing a nonzero spin current,$K_x$.  The maximal value of $\tau$ also implies that  in this phase, every spin is maximally entangled with its complement, \textit{i.e.} the rest of the system or in other words, entanglement is purely \textit{multispin} and the state of the system is a maximally entangled state, an $n$-qubit equivalent of two-qubit Bell states in this phase.   Further, we find that there is no factorizing curve in the transition from a magnetically ordered phase to the chiral phase.
 
 \subsection{\label{sec:Dz}Effect of a longitudinal DMI}
In  previous work~\cite{pradeepAIP, pradeepPRB}, we had shown that a longitudinal DMI $D_z$(which we interpreted there as an electric field ) in the spin $1/2$ anisotropic $XXZ$ model does not lead to any new phases; it only modifies the phase boundaries; increasing the disordered phase. For $D_z$ smaller than a certain critical strength $D_c$, there are two gapped phases: an antiferromagnetically ordered phase ($AFM_z$) for $h_x<h_{cr}$ and a gapped ferromagnetic ($FM_x$) phase for $h_x>h_{cr}$. The $AFM_z$ phase corresponds to a phase with a staggered magnetization $M^z_s$ along the $z$-direction and  a uniform magnetization $M^x$ along the $x$-direction~\cite{pradeepPRB}.  There is also a finite chiral current $\langle K^z \rangle$ (termed as electric polarization $P^y$ in Ref.~\cite{pradeepPRB}) in this phase. The transition to the $FM_x$ phase occurs at a critical transverse field strength $h_x=h_{cr}$, the value depending on the strength of $D_z$.  
 We discuss here the behaviour of the one-tangle and two spin concurrence in the presence of a longitudinal DMI. 
\begin{figure*}[!htpb]
\includegraphics[width=3.0in]{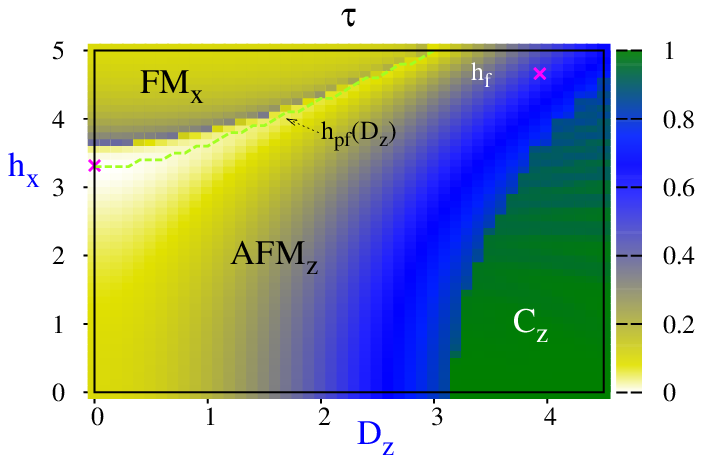}{(a)}
\includegraphics[width=3.0in]{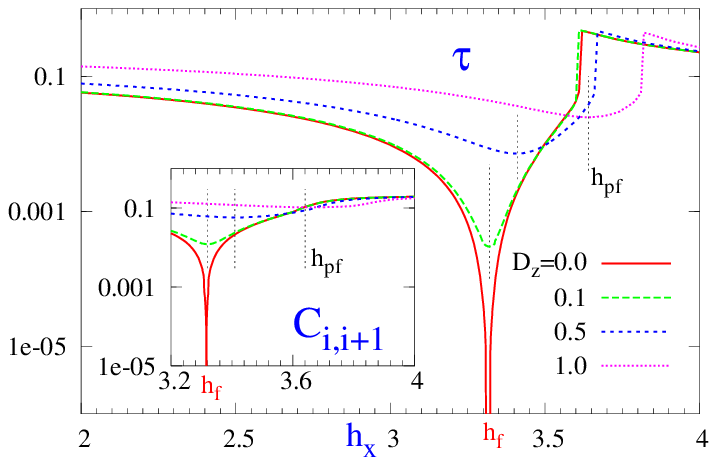}{(b)}
\includegraphics[width=3.0in]{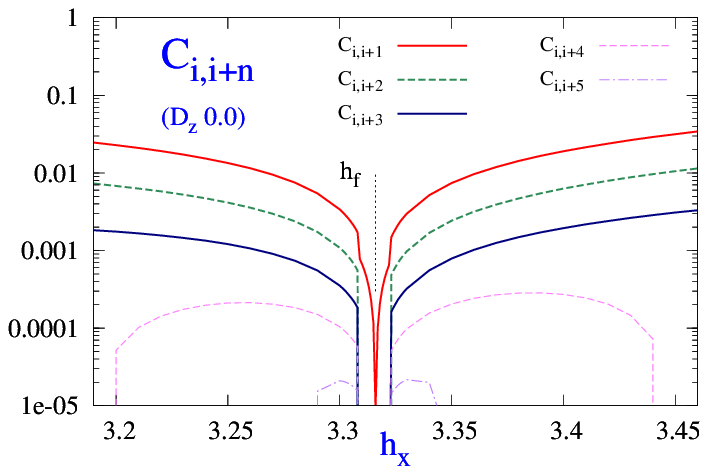}(c)
\includegraphics[width=3.0in]{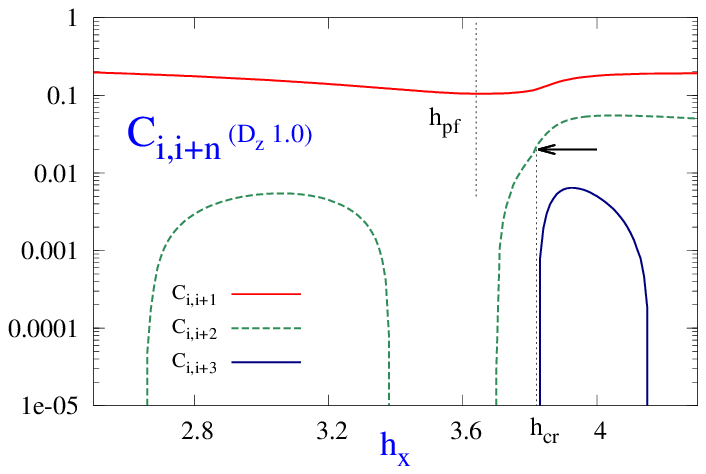}{(d)}
\caption{(a) The $h_x-D_z$ dependence of the one-tangle $\tau$. The factorising field, $h_f$, and the pseudofactorising (PF) field, $h_{pf}(D_z)$, both occurring inside the $AFM_z$ phase are marked in the figure.  The DMRG computations were performed on a 256 site chain with open boundary conditions. The anisotropy $\Delta $ is set to 
$4.5$.  The truncation error in the DMRG program was set to $10^{-6}$, with $10$ sweeps and maximum number of states kept not exceeding $70$. 
(b) The semilog plot of the one-tangle as a function of $h_x$ in absence and presence of $D_z$. It. vanishes at $h_x=h_f \sim 3.316$ for $D_z=0$, but attains a nonzero minimum value at $h_{pf}$ for $D_z=,0.1,0.5,1.0$, marked by vertical dotted lines from left to right respectively. The inset shows the similar $h_x-D_z$ dependence of the nearest-neighbour concurrence, $C_{i,i+1}$. 
 (c) Two spin  Concurrence, $C_{i,i+n}$, $n=1,2,3,4,5$, in the absence of $D_z$. All vanish at $h_x=h_f=3.316$. The range increases in the vicinity of $h_f$.
  (d) In the presence of $D_z$, the number of entangled pairs has decreased and $C_{i,i+1}$ is a finite minimum at $h_x=h_{pf}=3.64$. The kink in $C_{i,i+2}$ marked by the arrow indicates the $AFM_z-FM_x$ phase transition.
(In panels (b), (c) and (d), the computations were performed on a $N=64$ site chain. We also added a small  $h_z=0.1$ to break the $Z_2$ symmetry.) }
\label{fig:tau1_4p5d_hxE}
\end{figure*}
 
  In Fig.\ref{fig:tau1_4p5d_hxE}(a), we show the $h_x-D_z$ dependence of the one-tangle. The regions marked  ($AFM_z$)$FM_x$  correspond to gapped phases with (anti)ferromagnetic order along $(z) x$ direction respectively while the region marked $C_z$ corresponds to the gapless magnetically disordered chiral phase.  One can see from the figure  that in general, the one-tangle has a small value  in the magnetically ordered regions while it becomes large in the chiral phase.  In the absence of $D_z$, the one-tangle vanishes identically at the factorizing field,  $h_x =h_{f}$ which occurs inside the $AFM$ phase just before the QPT to the $FM$ phase. On the other hand, for a non-zero $D_z$, the one-tangle does not vanish at any $h_x$; however, there is a field $h_{pf}$ at which it goes to a (nonzero) minimum.  The $h_x$ dependence  of the tangle can be better seen from the semilog plot of $\tau$  as a function of $h_x$ shown in Fig. \ref{fig:tau1_4p5d_hxE}(b) for representative values of $D_z$.  The tangle shows a non-monotonic $h_x$ dependence both in the absence and presence of $D_z$; $\tau$ monotonically decreases with increasing $h_x$,  goes to a minimum at $h_x=h_f (=h_{pf})$ for $D_z=0( \neq 0)$, sharply rises near the critical field $h_x=h_{cr}$ and  then decreases monotonically with $h_x$.  The  NN concurrence shows  similar non-monotonic $h_x$ behaviour as that of the tangle. As can be seen from the semilog plot of the nearest neighbour (NN) concurrence $C_{i,i+1}$ shown in the inset of the figure.  We note that just like the one tangle, in the presence of $D_z$, the NN concurrence  does not vanish identically at any $h_x$ although it goes to a (non-zero) minimum value  at $h_x=h_{pf}$, unlike the case $D_z=0$ where the concurrence vanishes identically at $h=h_f$.  Thus, we find that both the one- tangle and NN concurrence go to a non-zero minimum at $h_x =h_{pf}$ for $ D_z \neq 0)$,  corroborating the result of a minimum non-zero entanglement in the ground state at $h_x=h_{pf}$ in the presence of $D_z$.

                We also examine the behaviour of the concurrence for spins  separated by distances $n>1$.  The log plots of the $h_x$-dependence of $C_{i,i+n}$ (for $D_z =0, 1$) are shown in panels (c,d) of  Fig. \ref{fig:tau1_4p5d_hxE}.  It can be seen from panel (c) that in the absence of $D_z$,  the concurrence $C_{i, i+n}$ vanishes for any separation of the spins at the factorizing field $h_f$. Furthermore, the range of pairwise entanglement diverges with more and more concurrences beyond nearest neighbours becoming non-zero on both sides of the factorizing field ($h_f \sim3.3$) or in other words,  there is an accumulation of pairwise entanglement about the factorizing  point with the concurrences all vanishing exactly at the  factorizing field. Such a divergence in the range of the concurrence is characteristic of an ET occuring at the factorizing field with a crossover from one type of entanglement to another ~\cite{fubini2006}.  (We also mention here that similar results are obtained for the higher order concurrence for a transverse DMI  ($D_x < D_{U_x}$), with the range of the two spin concurrence showing a diverging behaviour  similar to that shown in Fig.~\ref{fig:tau1_4p5d_hxE} (c).)  The behaviour  of the higher order concurrences in the presence of $D_z$ as shown in  panel (d) of  Fig.~\ref{fig:tau1_4p5d_hxE} is different. When  $D_z \neq 0$,  all the two spin  concurrences vanish identically at $h=h_{pf}$ for separations $n \geq 2$.  In the vicinity of $h_{pf}$, the range of the concurrence increases but remains finite, with the range decreasing as $D_z$ increases.

\section{\label{sec:ET} Entanglement transition in the presence of transverse and longitudinal DMI}
 
In this section, we discuss the nature or type of the entanglement in the different phases  and the existence of an entanglement transition in the presence of the DMI.  The type of the entanglement between two spins in the spin system can be probed by using the relations between the pairwise concurrence and the occupation probabilities relative to specific sets of two-spin quantum states~\cite{fubini2006}.  Specifically, we study the dependence of the occupation probabilities of different basis states on  the transverse magnetic field and DMI.  We also discuss the existence  of an ET by studying the nature of the two spin concurrence.  The diagonal elements of the reduced  two-spin density matrix are the respective occupation probabilities of the basis states.  The different bases considered for each pair of spins are:
\begin{align}
\mathcal{Z}_1 &\equiv \{|u_{I}\rangle,|u_{\text{II}}\rangle,|u_{\text{III}}\rangle,|u_{\text{IV}}\rangle\};  \quad \mathcal{X}_1 \equiv \{|v_{\text{I}}\rangle,|v_{\text{II}}\rangle,|v_{\text{III}}\rangle,|v_{\text{IV}}\rangle\} \nonumber \\ 
\mathcal{Z}_2 &\equiv \{|e_1\rangle,|e_2\rangle,|e_3\rangle,|e_4\rangle\}; \quad  \mathcal{X}_2 \equiv \{|f_1\rangle,|f_2\rangle,|f_3\rangle,|f_4\rangle\}\nonumber \\
\mathcal{Z}_3 &\equiv \{|u_{\text{I}}\rangle,|u_{\text{IV}}\rangle,|e_3\rangle,|e_4\rangle\}; \quad \mathcal{X}_3 \equiv \{|v_{\text{I}}\rangle,|v_{\text{IV}}\rangle,|f_3\rangle,|f_4\rangle\}
\end{align}
where,
\begin{align}
|u_{\text{I}}\rangle&\equiv|\uparrow\uparrow\rangle,|u_{\text{II}}\rangle\equiv|\uparrow\downarrow\rangle, 
|u_{\text{III}}\rangle\equiv|\downarrow\uparrow\rangle,|u_{\text{IV}}\rangle\equiv|\downarrow\downarrow\rangle,\\ \nonumber
|e_1\rangle &= \frac{1}{\sqrt2}(|u_{\text{I}}\rangle+|u_{\text{IV}}\rangle),|e_2\rangle = \frac{1}{\sqrt2}(|u_{\text{I}}\rangle-|u_{\text{IV}}\rangle),\\ \nonumber
|e_3\rangle &= \frac{1}{\sqrt2}(|u_{\text{II}}\rangle+|u_{\text{III}}\rangle),|e_4\rangle = \frac{1}{\sqrt2}(|u_{\text{II}}\rangle-|u_{\text{III}}\rangle)
\label{eq:prob_basis}
\end{align}
Here the single-spin states $|\uparrow\rangle|$ and $|\downarrow\rangle$ are eigenstates of the operator $S_z$ with eigenvalues $1/2$ and $-1/2$ respectively. 
$\mathcal{Z}_1$ and $\mathcal{Z}_2$ are the \textit{standard} and \textit{Bell} bases respectively, and $\mathcal{Z}_3$ is called the \textit{mixed} basis~\cite{fubini2006}.
The two-spin bases $\mathcal{X}_1, \mathcal{X}_2, \mathcal{X}_3$ are the \textit{standard},  \textit{Bell} and \textit{mixed} basis respectively, defined in terms of single-spin eigenstates of the operator $S_x$.
Specifically, 
\begin{align}
|v_{\text{I}}\rangle&\equiv|\rightarrow\rightarrow\rangle,|v_{\text{II}}\rangle\equiv|\rightarrow\leftarrow\rangle,
|v_{\text{III}}\rangle\equiv|\leftarrow\rightarrow\rangle,|v_{\text{IV}}\rangle\equiv|\leftarrow\leftarrow\rangle,\\ \nonumber
|f_1\rangle &= \frac{1}{\sqrt2}(|v_{\text{I}}\rangle+|v_{\text{IV}}\rangle),|f_2\rangle = \frac{1}{\sqrt2}(|v_{\text{I}}\rangle-|v_{\text{IV}}\rangle),\\ \nonumber
|f_3\rangle &= \frac{1}{\sqrt2}(|v_{\text{II}}\rangle+|v_{\text{III}}\rangle),|f_4\rangle = \frac{1}{\sqrt2}(|v_{\text{II}}\rangle-|v_{\text{III}}\rangle)
\end{align}
and the single-spin states $|\rightarrow\rangle|$ and $|\leftarrow\rangle$ are eigenstates of the operator $S_x$ with eigenvalues $1/2$ and $-1/2$ respectively.
The states $|\rightarrow\rangle|$ and $|\leftarrow\rangle$ may be written in terms of the single-spin eigenstates of the operator $S_z$ as: 
\begin{align}
|\rightarrow\rangle| &= \frac{1}{\sqrt2}(|\uparrow\rangle+|\downarrow\rangle), |\leftarrow\rangle = \frac{1}{\sqrt2}(|\uparrow\rangle-|\downarrow\rangle)
\end{align}
which leads to the  identities:
\begin{align}
|f_1\rangle &= |e_1\rangle, |f_2\rangle = |e_3\rangle, |f_3\rangle = |e_2\rangle,  |f_4\rangle = |e_4\rangle
\end{align}

Denoting the probabilities of the states in the Bell basis $\mathcal{Z}_2$ as $p_i$, we can see from Eq. 8 that $p_1, p_2$ correspond to states with parallel spins while $p_3, p_4$ correspond to states with antiparallel spins.  In all the three bases, the states with parallel and antiparallel spins do not mix with each other.
Due to the above identity, the probabilities $p_i$ of the states in the Bell basis, $\mathcal{Z}_2$  can be related to the probabilities $p_{ix}$ of the states in the Bell basis $\mathcal{X}_2$ as:
\begin{align}
p_{1x} &= p_1; \quad  p_{2x} = p_{3};\quad  p_{3x} = p_2;\quad  p_{4x} = p_4
\label{xz_basis}
\end{align}

\begin{figure}[!htpb]
\includegraphics[width=3.0in]{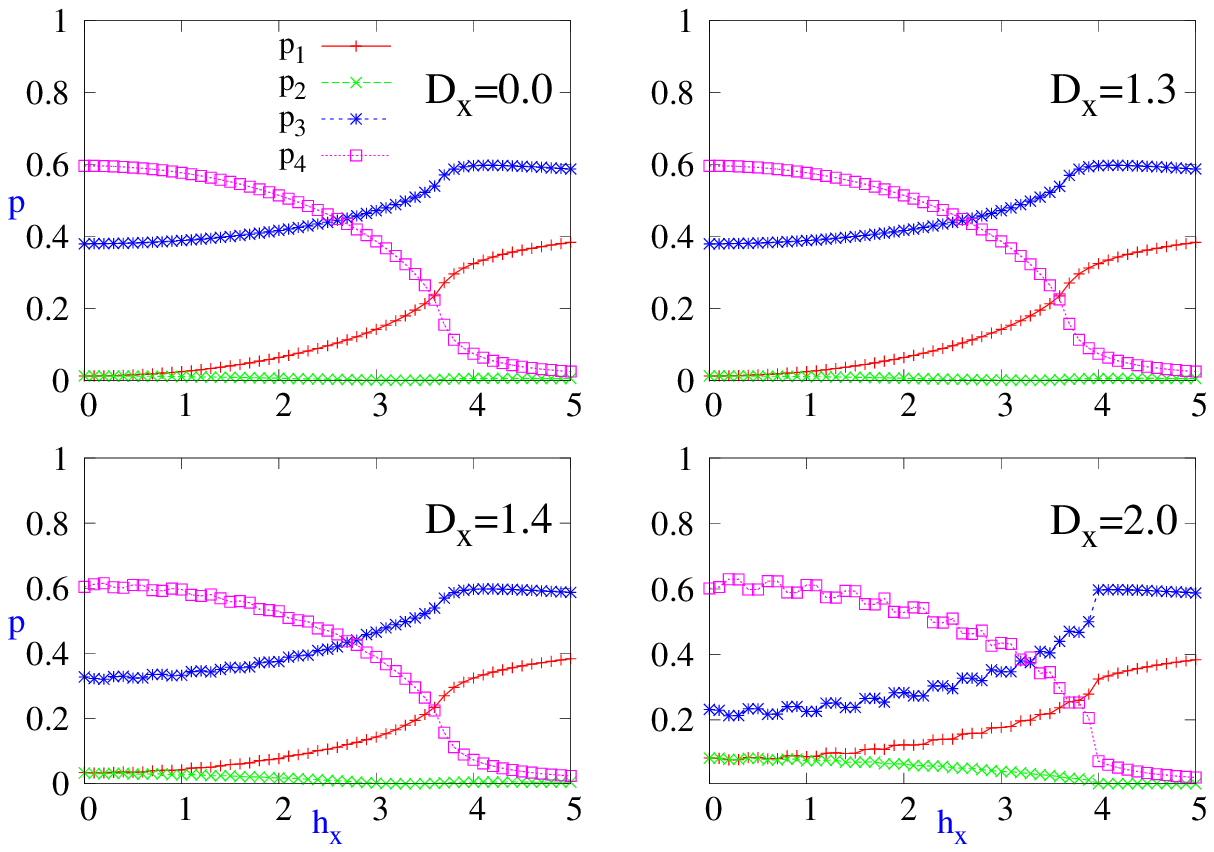}{(a)}
\includegraphics[width=3.0in]{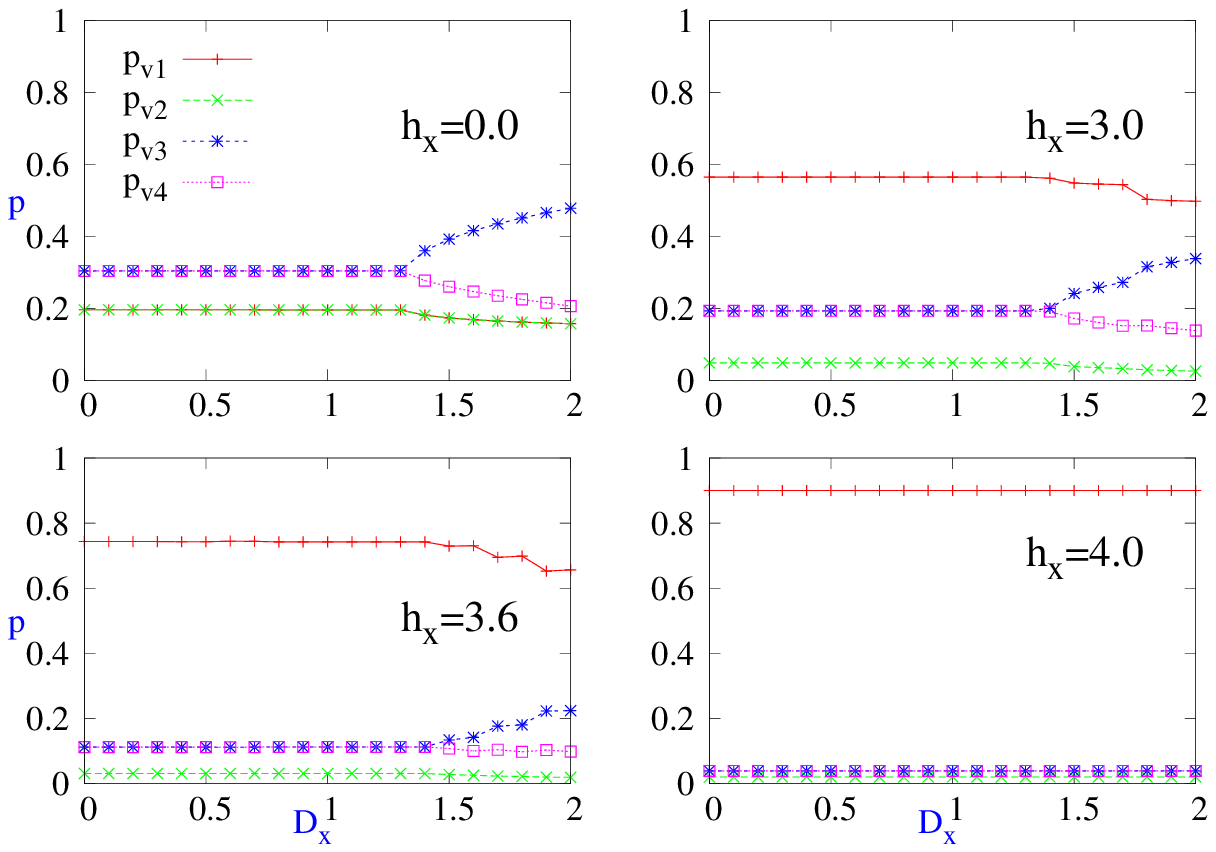}{(b)}
\caption{(a)The $h_x$ dependence of  the occupation probabilities of two spin state in the Bell basis for the central bond of a $64$ site chain for diffferent $D_x$ values. (b)  The $D_x$ dependence of  the occupation probabilities of two spin states in the chiral $K^x$  basis for the central bond of a $64$ site chain for diffferent $h_x$ values. All other parameters are as in Fig. 1.
}
\label{fig:SySz_probs}
\end{figure}

Fig. \ref{fig:SySz_probs}(a) shows the occupation probabilities in  the Bell basis $\mathcal{Z}_2$ for the central bond of the chain as functions of $h_x$ for different values of $D_x$.  
In the absence of DMI ($D_x=0$), we can see from the plot that in the $AFM_z$ phase ($h_x<h_{cr}$),  $p_3$ and $p_4$ which are the probabilities corresponding to the antiparallel  spin states dominate over the probabilities, $p_1$ and $p_2$; also  $p_4 >p_3$.  For $h_x > h_{cr}$,  i.e, in the $FM_x$ phase, it is convenient to use the $\mathcal{X}_2$ basis for interpreting the plot: then we see that  $p_3 (= p_{2x})$ is the dominant probability followed by $p_1(= p_{1x})$ while  $p_2(= p_{3x})$ is the smallest occupation probability.  As $h_x$ increases, $p_4(= p_{4x})$ decreases monotonically going to zero asymptotically in the limit $h_x \rightarrow \infty$.  Thus, for $h_x>h_{cr}$, the probabilities, $p_{1x}$ and $p_{2x}$, of the parallel spin states (with respect to the Bell basis $\mathcal{X}_2$) are the largest and 
$p_{2x} >p_{1x}$.  These observations indicate the presence of antiparallel entanglement in the $z$-component of the spin on the AFM side of the factorizing field and parallel entanglement in the $x$-component on the other side. We also note  that the transverse magnetic field breaks the symmetry in probability between the states $e_1$ and $e_2$;  in the absence of $h_x$, $p_1=p_2$,  a finite $h_x$ favours the occurrence of the state $e_1$ over the state $e_2$.   In the presence of the transverse DMI, the occupation probabilities show similar behaviour for $D_x<D_c$. 

  In order to understand better the change in the probabilities across the $AFM_z$ -chiral phases as well as the $FM_x$ - chiral phases,  the appropriate basis to consider in the chiral phase are the eigenstates of the chiral  operator $K^x=S_i^y S_{i+1}^z - S_i^z S_{i+1}^y$.   Expressing the Pauli matrices in the representation:
$
|\uparrow\rangle =  
\begin{pmatrix}
1 & 0 
\end{pmatrix}^T; \quad
|\downarrow\rangle =  
\begin{pmatrix}
0 & 1
\end{pmatrix}^T
$, 
 $K^x$ may be written as:
\begin{eqnarray}
K^x = S^y_i S^z_{i+1} - S^z_i S^y_{i+1}=
i
\begin{pmatrix}
0 & 1 & -1 & 0\\
-1 & 0 & 0 & 1\\
1 & 0 & 0 & -1\\
0 & -1 & 1 & 0
\end{pmatrix}
\end{eqnarray}
The eigenvalues of this operator are $0$, $0$, $-2$, and $2$ and the corresponding eigenvectors are:
\begin{eqnarray}
v_1=0.5\begin{pmatrix}
1 & 1 & 1 & 1
\end{pmatrix}^T,
v_2=0.5\begin{pmatrix}
1 & -1 & -1 & 1
\end{pmatrix}^T,\nonumber \\
v_3=0.5\begin{pmatrix}
1 & -i & i & -1
\end{pmatrix}^T,
v_4=0.5\begin{pmatrix}
1 & i & -i & -1
\end{pmatrix}^T
\end{eqnarray}
$v_1=|\rightarrow \rightarrow \rangle$ and $v_2=|\leftarrow \leftarrow \rangle$ are eigenstates of the $S^x_i \otimes S^x_{i+1}$ operator and are product states, hence unentangled. On the other hand, $v_3$ and $v_4$ are entangled and cannot be written as a separable product of two single-spin states. Denoting the corresponding probabilities as $p_{v1}$, $p_{v2}$, $p_{v3}$, and $p_{v4}$, we note that $p_{v1}=p_I$, the probability for the state $|\uparrow \uparrow \rangle$ and $p_{v2}=p_{IV}$, the probability for the state $|\downarrow \downarrow \rangle$, which are both non-entangled states.  We show the $D_x$-dependence of the probabilities of the $K^x$-basis for the central bond of the chain for different values of $h_x$ in Fig. \ref{fig:SySz_probs}(b).  The first thing we note is that in $AFM_z$ and $FM_x$ phases, all probabilities are independent of $D_x$;  the values are  determined only by $h_x$ while in the chiral phase, the probabilities show both $h_x$ and $D_x$ dependence.
In the $AFM_z$ phase, $p_{v3}=p_{v4}$.  As expected, increase in $h_x$ increases $p_{v1}(=p_I)$ towards saturation and decreases $p_{v2}=p_{IV})$ towards zero.  In the $FM_x$ phase, ($h_x\geq h_{cr}$),  $p_{v1}$ is maximum ($\sim1$) and $p_{v2}\sim0$. The other two probabilities  $p_{v3}$ and $p_{v4}$ are both very small ($\sim0$) in the ferromagnetic phase.  There is a sharp change in behaviour of $p_{v1}$ at the chiral-$FM_x$ QPT (for $D_x  \approx  D_c$),   $p_{v3}$ and $p_{v4}$ which are equal in the $AFM_z$ phases split in the chiral phase, where $p_{v3}>p_{v4}$.  As $D_x$ increases further,  the difference between $p_{v3}$ and $p_{v4}$ increases with $p_{v3}$ saturating towards the value $1/2$. \\

\begin{figure}[!htpb]
\includegraphics[width=3in]{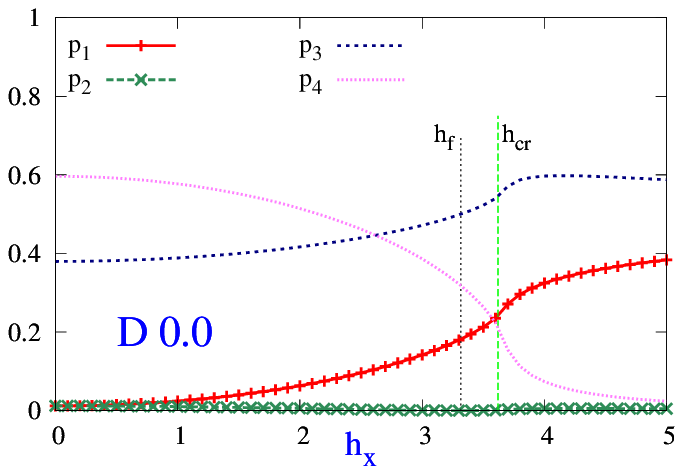}{(a)}
\includegraphics[width=3in]{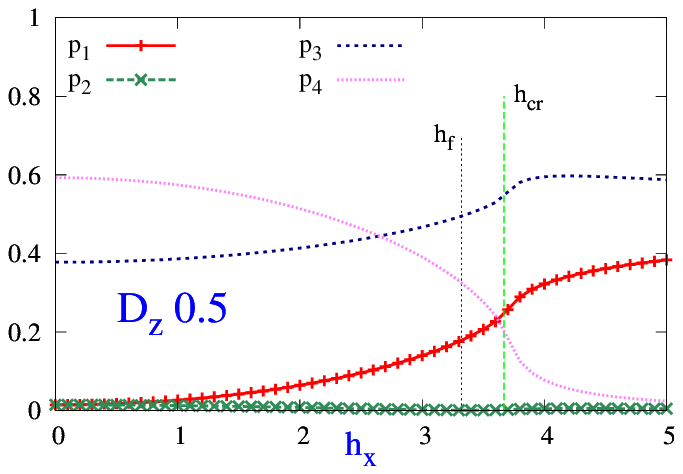}{(b)}
\includegraphics[width=3in]{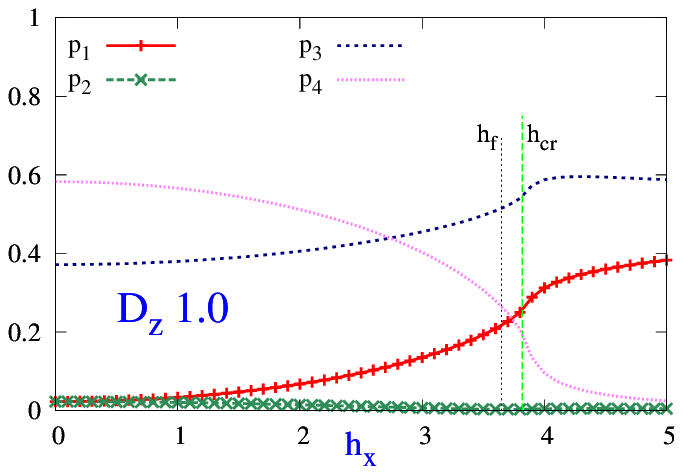}{(c)}
\caption{(a) Occupation probabilities of two spin states in the Bell basis, $\mathcal{Z}_2$, as functions of $h_x$, for (a) $D_z=0.0$, (b) $D_z=0.5$, (c) $D_z =1.0$ for a $N=64$ site chain.
A small $h_z=0.1$ has been added to prevent degeneracy effects due to $Z_2$ symmetry. All other parameters are as in Fig. 1.}
\label{fig:Dz_probs_hx}
\end{figure}

We now consider the case of a longitudinal DMI.  We show the $h_x$ dependence of the probabilities in the Bell basis  in the presence of a longitudinal DMI $D_z$ in  Fig. \ref{fig:Dz_probs_hx}(a-c).  We observe that even in the presence of $D_z$,  the antiparallel spin states of the $\mathcal{Z}_2$ basis (Bell basis for the $S^z$-components) dominate in the $AFM$ phase while the parallel spin states of the $\mathcal{X}_2$ basis (Bell basis for the $S^x$-components) dominate in the $FM_x$ phase.
(We do not show the results for large $D_z$ values,  since the results are similar to the case of a transverse DMI with the difference that the chiral phase here is characterized by the chiral current $K_z$.)

\begin{figure}[!htpb]
\includegraphics[width=3.0in]{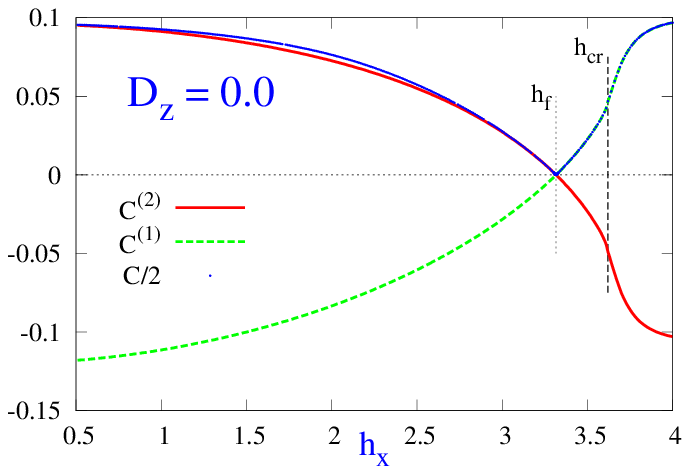}{(a)}
\includegraphics[width=3.0in]{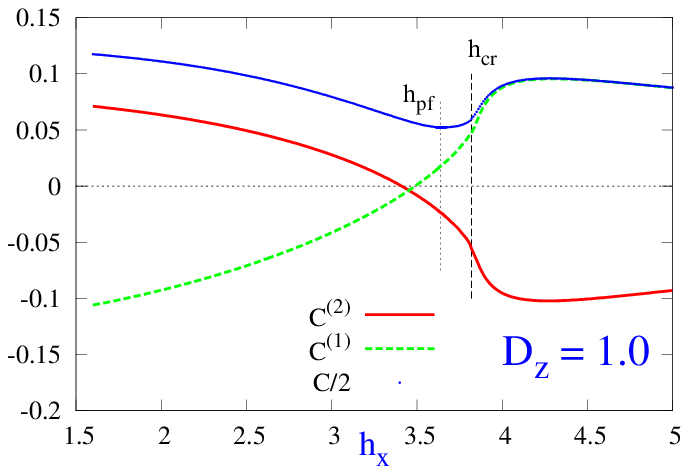}{(b)}
\caption{Comparison of the analytic expressions for the concurrence $C_{ij}^{(1)}$ and $C_{ij}^{(2)}$ defined in Eq. \ref{eq:pdn_C}, with the numerically calculated half of the concurrence, $C/2$, (a) in the absence of $D_z$, and (b) in presence of $D_z$.}
\label{fig:C1_C2_pdn}
\end{figure}

    The existence of an entanglement transition  across the $AFM_z -FM_x$ transition can be obtained by studying the nature of the two-spin concurrence~\cite{roscilde2004, amicoRange2006}.  In the absence of spontaneous symmetry breaking  ($M^z =0$), the  concurrence can be expressed as~\citep{amicoRange2006, roscilde2004}:
$\frac{C_{ij}}{2}= \text{max}\{0, C_{ij}^{(1)}, C_{ij}^{(2)}\}$
where  
\begin{eqnarray}
 C_{ij}^{(1)} = \sqrt{ (g_{ij}^{zz} - g_{ij}^{yy})^2 + (g_{ij}^{zy} + g_{ij}^{yz})^2 }-\sqrt{(\frac{1}{4} - g_{ij}^{xx})^2 - \delta S_x^2} \nonumber \\
 C_{ij}^{(2)} = \sqrt{ (g_{ij}^{zz} +g_{ij}^{yy})^2 + (g_{ij}^{zy} - g_{ij}^{yz})^2 }-\sqrt{(\frac{1}{4} + g_{ij}^{xx})^2 - M_x^2} \nonumber \\
\label{eq:pdn_C}
\end{eqnarray}
 $C^{(2)}$ represents antiparallel entanglement (along $z$ direction ) while $C^{(1)}$ denotes parallel entanglement along $x$ direction.
In Fig. \ref{fig:C1_C2_pdn}, we show the $h_x$ dependence of  $C_{ij}^{(1)}, C_{ij}^{(2)}$ and the numerically computed $\frac{C_{ij}}{2}$ for the nearest neighbour spins. 
We can see from panel (a), that  in the absence of $DMI$,  the factorizing field $h_x=h_f$ distinguishes two field regions with different expression for the concurrence:
$C_{ij}^{(1)} <0<C_{ij}^{(2)}$  for $h_x<h_f$ whereas  $C_{ij}^{(2)} <0<C_{ij}^{(1)}$ for $h_x >h_f$.
$C_{ij}^{(1)}$ and $C_{ij}^{(2)}$  vanish and cross exactly at $h_x=h_f=3.316$. Thus there is an entanglement transition across the factorizing field with the nature of the entanglement changing from an antiparallel entanglement along $z$ direction to parallel entanglement along $x$ direction.
We also note that it can be seen from our plot that  for $h_x>h_f$ the concurrence estimated from $C^{(1)}$ matches very well with the numerically computed concurrence but for $h_x<h_f$,  the matching between the concurrence estimated from $C^{(2)}$ and the numerically evaluated value is not that accurate.
This is in agreement with earlier results that while in the presence of spontaneous symmetry breaking $M^z \neq 0$ as happens when $h_x<h_{cr}$, the concurrence estimated from Eq.~\ref{eq:pdn_C} are not in general expected to match with the actual values of the concurrence;  the concurrence estimated from the above equation is accurate when $h_x>h_f$ but represents a lower bound to the concurrence for $h_x<h_f$.  A similar plot is obtained for a transverse DMI $D_x (<D_{U_x})$.

In the presence of the longitudinal DMI $D_z$ as shown in Fig. \ref{fig:C1_C2_pdn}(b),  $C^{(2)}$ and $C^{(1)}$ do not cross at $h_{pf}$. Further, we can see that $C_{ij}^{(2)}$ is a very poor approximation for $C/2$ for $h_x<h_{pf}$ although $C^{(1)}$ matches very well with $C$ for $h>h_{cr}$.
However, one observes that the pseudofactorizing field still distinguishes between regions with a dominant $C_{ij}^{(2)}$ for  $h_x<h_{pf}$ and a dominant $C^{(1)}$ when $h_x > h_{pf}$ indicating the presence of an entanglement transition even in the presence of the longitudinal DMI.  This indicates that although there is no factorizing field in the presence of a longitudinal DMI,  at which the ground state is fully classical, there is still an entanglement transition across the pseudo-factorizing field with the nature of the entanglement changing across the QPT.  As seen from the results in the previous section, in the vicinity of $h_{pf}$, there is an enhanced but finite  range of the concurrence in contrast to the diverging range of the concurrence associated with the ET across the factorizing field in the absence of $D_z$.

\section{\label{sec:coherence}Quantum coherence and asymmetry}

  We discuss, in this section, the connections between  the factorizability phenomenon in the $AFM_z$ phase  with symmetry and coherence properties. The factorizability condition for the ground state $|G \rangle$, namely, $|G\rangle$ is factorized if and only if the single spin tangle vanishes for all spins $k$ in the lattice can be recast as a condition for the  invariance of the state under unitary rotations about the local magnetization axis $\hat m_k \equiv\vec M_k/|\vec M|$ at the site $k$ ~\cite{giampaolo2009}: 
 $$|G \rangle = \bar U_k |G\rangle;\quad \bar U_k =  \otimes _{j\neq k} I_j \otimes \bar A_k;\quad \bar A_k = \hat m_k \cdot {\vec S}_k$$.
The trace distance between the states $|G\rangle$ and $\bar U_k |G \rangle$ is related to the one tangle as $d_k \equiv \sqrt{1 - |\langle G \bar U_k G\rangle |^2}  = \sqrt \tau_k$.  Thus the one-tangle can also be interpreted as a quantifier of the rotation asymmetry of the $N$ spin ground state about the local magnetization axis at each lattice site. 
The vanishing of the one-tangle in the $AFM_z$ phase,  exactly at the factorizing field implies then that  the ground state of the  $XXZ$ model acquires a  $U(1)$ rotation symmetry  about the  local magnetization axis  $\hat m_k$,  although the Hamiltonian itself does not commute with $\bar U_k$.   An additional  transverse DMI preserves the factorizability phenomenon in the $AFM_z$ phase and hence the  $U(1)$ rotation symmetry at  $h_x=h_f$.  On the other hand,  since in the presence of a longitudinal DMI, the one spin tangle does not vanish at any field, it implies that the $U(1)$ rotation symmetry about the local magnetization axis at each site is broken at all values of $h_x$ with minimal breaking of the symmetry at $h_x=h_{pf}$.  

  Another natural information theoretic measure to analyze the asymmetry properties of a quantum state is the Wigner-Yanase skew information(WYSI), $I(\rho, X) = -\frac{1}{2} Tr[\sqrt \rho, X]^2$ ~\cite{wigner1963, girolami2014}.  Here $\rho$ is a quantum state(in general, mixed) and $X$ is a physical observable. A state $\rho$ is left invariant by measuring an observable $X$ (assumed bounded and non-degenerate) if and only if it does not show coherence in the $X$ eigenbasis, being a mixture of eigenstates of the observable, i.e.. $[\rho, X] =0$.   For a bipartite composite state $\rho_{AB}$, we can define the local quantum coherence or asymmetry  with respect to the first subsystem as $I(\rho_{AB}, X_A \otimes I_B)$.  The skew information $I(\rho, X)$ is related to the uncertainty of measuring  the observable $X_A $ with respect to the composite state.  For a pure state $\rho$, it is the variance~\cite{luoprl2013} $I(\rho, X) = V(X) =  Tr (X^2 \rho) - (Tr X \rho)^2$.  
  
\begin{figure}
\includegraphics[width=3.0in]{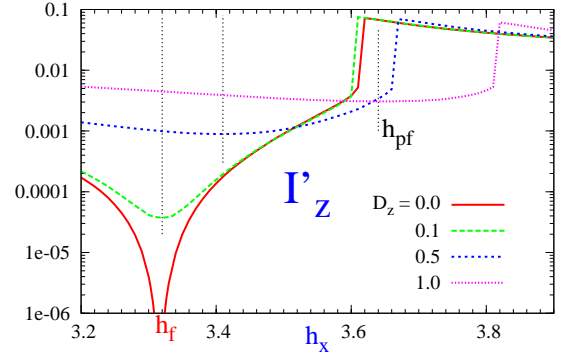}{(a)}
\includegraphics[width=3.0in]{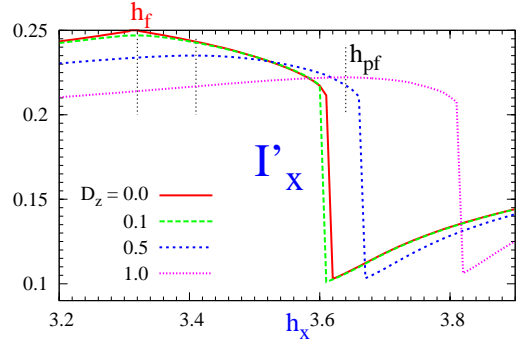}{(b)}
\includegraphics[width=3.0in]{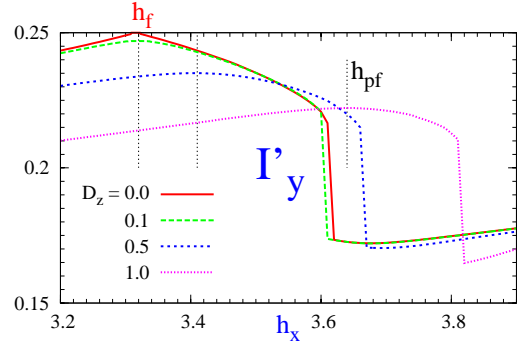}{(c)}
\caption{The $h_x$  dependence of the local quantum coherences (a) $I^{\prime}_z= \mathcal{I}(\rho,S^{\prime z}_i \otimes I_{i+1})$, (b)$I^{\prime}_x =\mathcal{I}(\rho,S^y_i\otimes I_j)$, and (c)$I^{\prime}_y=\mathcal{I}(\rho,S^y_i \otimes I_j)$ for representative values of $D_z$. 
We set the following parameters: $h_z=0.1$, $N=64$, and $\Delta=4.5$. The label in red, $h_f$, in each panel points to the kink occurring at the factorizing field, $h_f$ for $D_z=0$. The black dotted lines mark the position of the pseudofactorizing field $h_{pf}$, for  respective $D_z$ values . }
\label{fig:LQC_2aa}
\end{figure}

   We compute WYSI in a rotated spin basis chosen at each site,  with the new $z$-axis pointing along  the local magnetization axis and the other axes lying in the plane perpendicular to it:
 \begin{eqnarray}
S^{\prime z}  = \sin \bar \theta \cos \bar \phi S^x + \sin \bar \theta \sin \bar  \phi S^y + \cos \bar \theta S^z  \nonumber \\
S^{\prime x} = \cos  \bar \theta \cos \bar \phi S^x + \cos \bar \theta \sin \bar \phi S^y -\sin \bar \theta S^z \nonumber \\
S^{\prime y} = -\sin \bar \phi S^x + \cos \bar \phi S^ y
\end{eqnarray}
and the new coordinate axes have been defined as 
\begin{eqnarray}
 \hat z^{\prime}_k \equiv  \hat m_k = (\sin \bar \theta_k  \cos \bar \phi _k,  \sin \bar \theta_k,  \cos \bar \theta_k); \nonumber \\
  \hat x^{\prime}_k = (\cos \bar \theta_k \cos \bar \phi_k, \cos  \bar \theta_k \sin \bar \phi_k, - \sin \bar \theta_k); \nonumber \\
   \hat y^{\prime}_k =( -\sin \bar \phi_k, \cos \bar \phi_k, 0)
 \end{eqnarray} 
   We begin by discussing the skew information in the many body ground state with respect to $X_k^z = (\prod_{j \neq k} \otimes I_j)  \otimes S^{\prime z}_k$. Since the ground state is a pure state, the skew information in the ground state is the variance $V(X_k^z) = \frac{1}{4} - (\vec M_k)^2 = \frac{1}{4} \tau_k $.  Thus  $I(\rho_G, X_k^z) = \tau/4 $(we drop the site index due to the translation invariance) which confirms that the one-tangle  quantifies the $U(1)$ rotation asymmetry (about the  magnetization axis) in the ground state.  Consider now the skew information contained in the two spin reduced density matrix $\rho^{(2)}_{i, i+1}$.  The $h_x$ dependence of the quantum coherences $I(\rho^{(2)}_{i,i+1}, S^{\prime a}_i\otimes I_{i+1})$, for $a=x,y,z$,  computed in the rotated spin basis (denoted as $I^{\prime}_a, a=x,y,z$) for representative values of $D_z$   are shown  in Fig.~\ref{fig:LQC_2aa}. (In the presence of a transverse DMI, one obtains the same plots as in the absence of $DMI$). 
 In all three panels (a-c), we observe common features, namely, \\
\noindent (i) a sharp discontinuity in these coherence estimators (near-vertical drops in the plots) in the vicinity of the QPT from the $AFM_z$ to the $FM_x$ phase.\\
\noindent (ii)  a kink in the plots for all the WYSIs in the absence of DMI( $D_z=0$). This kink occurs at the factorizing field, $h_f(h_f\sim3.316)$. 
The longitudinal DMI  $D_z$ smoothens the kink. Such a smoothening of the kink near the QPT was observed earlier for the correlation functions~\cite{pradeepPRB}.\\
\noindent (iii)$I^{\prime}_x =I^{\prime}_y$ for $h_x<h_{cr}$.
In the absence of DMI,  and at the factorizing field, $I^{\prime}_z$ vanishes identically while $I^{\prime}_x(=I^{\prime}_y)$ takes the maximal coherent value $1/4$, signalling the  $U(1)$ rotation symmetry about the local magnetization axis $\hat m_k$.\\
Similar plots are obtained in the presence of a transverse $D_x (<D_c)$ which again confirm that a transverse DMI preserves the factorizability property and hence the local U(1) rotation symmetry
at the factorizing field.   On the other hand, for a non-zero $D_z$,  it can be observed from the plots that $ I^{\prime}_z$ does not vanish at any $h_x$; $I^{\prime}_{x(y)}$ also do not attain the maximal value at any $h_x$, indicating the violation of the local $U(1)$ symmetry.  The minimal deviation from these values are seen to occur at $h_x=h_{pf}$, with  $I^{\prime}_z$  acquiring its minimal (non-zero) value and  $I^{\prime}_{x(y)}$ acquiring their maximal value ($<1/4$)  at  $h_x=h_{pf}$.  The field $h_x=h_{pf}$ is thus the field at which the $U(1)$ rotation symmetry about the local magnetization axis is minimally broken.   

         \section{\label{sec:concl} Conclusions and Discussions}
In conclusion,  we have studied the effect of a longitudinal and transverse DMI on the quantum correlations present in the ground state of the anisotropic spin- $1/2$ $XXZ$ model in a transverse magnetic field. We have focussed on the factorizability and coherence properties which we have studied by computing  bipartite entanglement and coherence estimators like the one-tangle, two spin concurrence and  WYSI.  Our main result is that in the AFM phase, a transverse DMI preserves the factorizability property while longitudinal DMI destroys it.  
In the latter case, there is a `pseudofactorizing' field $h_{pf}$ at which the entanglement quantified by the one-tangle and nearest neighbour two-spin concurrence is finite but minimal.  Exactly at $h_{pf}$,  all  higher order two- spin concurrences $C_{i,j}, j >i+1$ vanish.  The existence of the pseudo-factorizing field can be associated with an entanglement transition with a finite increase in the range of the nearest neighbour two spin-concurrence in its vicinity. This is in contrast to the diverging  range of the nearest neighbour two spin concurrence in the vicinity of the factorizing field (in the absence of DMI).  We associate the existence or non-existence of the factorization phenomenon  with the absence  or presence of the chiral current: a transverse DMI does not lead to any chiral current in the $AFM_z$ phase while there is a finite chiral current in the presence of a longitudinal DMI.
  Specifically, in the presence of a longitudinal $D_z$,  the  non-factorizablity  phenomenon at any transverse magnetic field is related to  the violation of a $U(1)$ rotation symmetry about the local magnetization axis which physically manifests in a non-zero finite chiral current $\langle K^z\rangle $  in the $AFM_z$ phase~\cite{pradeepPRB}.  The asymmetry properties have been  quantified through the WYSI. The symmetry is broken minimally at  the pseudo-factorizing field.
        
             We can also relate the asymmetry to `frameness'  since an asymmetry of a state determines its  ability to be used as a reference frame for some measurement or to act as a reference frame under a superselection rule~\cite{bartlett, ahmadi2013, girolami2014}.  The asymmetry here  is the quantum coherence lost by applying the phase shift with respect to the eigenbasis of a `supercharge' $Q$~\cite{girolami2014}.  In our model, the number operator defined at each site as $ n_k \equiv S_k^z +1/2$ serves as the supercharge.  In the absence of $DMI$ ($D_z=0$), the factorized ground state  is specified completely by the local magnetization. Since all the unitary operations that are $U(1)$ invariant have the effect of merely changing the relative phases between the states $\psi_k$ at each site, this implies that the local magnetization axis at each site serves to specify locally the $z$ axis at each site. In a translationally invariant system, then this implies that the single site magnetization axis serves to specify the common $z$-axis  for the full system. However, due to the $U(1)$ rotation symmetry about the magnetization axis, what is lacking to specify a full Cartesian reference, is the angle between the local $x$ axis at different sites. This lack of a phase reference is a form of `decoherence'.  Essentially, the spin at each site points along the $z$ axis;  there is a number conservation, but there is complete loss of information about the relative phase between the  the spin states at neighbouring sites. The breaking of the $U(1)$ symmetry by the longitudinal DMI $D_z$ serves to specify the relative phase between neighbouring spin states or equivalently the relative angle between the $x$ axis at neighbouring sites.  Physically, the non-zero finite chiral current $\langle K^z\rangle $ serves as the macroscopic quantity to determine the phase relations between the spin states at neighbouring sites.   At $h_x=h_{pf}$, the symmetry is broken minimally; our numerical results indicate that the two -spin density matrix is an $X$-state leading to a phase coherent ground state with phase coherence between nearest neighbour spin states.   Akin to what happens in a superconductor with the breaking of $U(1)$ symmetry,  the ground state in the presence of a longitudinal DMI  can be described at the `pseudofactorizing' field by a phase coherent macroscopic quantum wave function with minimum non-zero bipartite entanglement.  Exactly at $h_{pf}$, then the local magnetization and chiral current are sufficient to specify the ground state completely; they also serve to specify the full  Cartesian reference frame for the many body system with the  chiral current serving as the macroscopic quantity to determine the phase reference. 

\acknowledgments{PD thanks SERB, DST, India for financial support through research grant. PT thanks UGC, India for financial support though the UGC-BSR Fellowship.}
\bibliography{manuscript_bib}

\end{document}